\documentclass[sigconf]{acmart}
\AtBeginDocument{%
  }

\copyrightyear{2026}
\acmYear{2026}
\setcopyright{cc}
\setcctype{by}
\acmConference[CIKM '26]{Proceedings of the 35th ACM International Conference on Information and Knowledge Management}{November 07--11, 2026}{Rome, Italy}
\acmBooktitle{Proceedings of the 35th ACM International Conference on Information and Knowledge Management (CIKM '26), November 07--11, 2026, Rome, Italy}
\acmDOI{10.1145/3799682.3841105}
\acmISBN{979-8-4007-2539-5/2026/11}
\usepackage{CJK}

\usepackage{paralist}
\usepackage{enumitem}
\usepackage{pifont}
\usepackage{makecell}
\usepackage{color}
\usepackage{xcolor}
\usepackage{mathrsfs}
\usepackage{mathtools}
\usepackage[ruled, linesnumbered]{algorithm2e}
\usepackage{multirow}
\usepackage{subfigure}  
\usepackage{subcaption}
\usepackage{float}
\usepackage{wrapfig}
\usepackage{subfloat}
\usepackage{caption}
\usepackage{adjustbox}
\usepackage{graphicx}
\newcommand{\Rmnum}[1]{\expandafter\@slowromancap\romannumeral #1@} 
\newcommand{\MethodName}{Rich-Media Re-ranker}
\usepackage{arydshln} 
\usepackage{multirow}
\usepackage{booktabs} 
\usepackage{siunitx}  
\usepackage[most]{tcolorbox}
\tcbuselibrary{listings}
\usepackage{fancyvrb} 
\tcbset{
  fontupper=\fontsize{8pt}{9pt}\selectfont,
  fonttitle=\bfseries\fontsize{8pt}{8pt}\selectfont
}
       
\begin{document}

\newlength{\equalcolwidth}


\title{Rich-Media Re-Ranker: A User Satisfaction-Driven LLM Re-ranking Framework for Rich-Media Search}

\author{Zihao Guo}
\affiliation{%
  \institution{SKLCCSE, School of Computer Science and Engineering\\Beihang University}
  \city{Beijing}
  \country{China}
}
\email{guozh@buaa.edu.cn}

\author{Ligang Zhou}
\authornote{Corresponding author.}
\affiliation{%
  \institution{Baidu Inc}
  \city{Beijing}
  \country{China}
}
\email{zhouligang@baidu.com}

\author{Zeyang Tang}
\affiliation{%
  \institution{Baidu Inc}
  \city{Beijing}
  \country{China}
}
\email{tangzeyang@baidu.com}

\author{Feicheng Li}
\affiliation{%
  \institution{Baidu Inc}
  \city{Beijing}
  \country{China}
}
\email{lifeicheng@baidu.com}

\author{Ying Nie}
\affiliation{%
  \institution{Baidu Inc}
  \city{Beijing}
  \country{China}
}
\email{nieying02@baidu.com}

\author{Zhiming Peng}
\affiliation{%
  \institution{Baidu Inc}
  \city{Beijing}
  \country{China}
}
\email{pengzhiming01@baidu.com}

\author{Qingyun Sun}
\authornotemark[1]
\affiliation{%
  \institution{SKLCCSE, School of Computer Science and Engineering\\Beihang University}
  \city{Beijing}
  \country{China}
}
\email{sunqy@buaa.edu.cn}

\author{Jianxin Li}
\affiliation{%
  \institution{SKLCCSE, School of Computer Science and Engineerin\\Beihang University}
  \city{Beijing}
  \country{China}
}
\email{lijx@buaa.edu.cn}

\renewcommand{\shortauthors}{Zihao Guo et al.}
\begin{abstract}
Re-ranking plays a crucial role in modern information search systems by refining the ranking of initial search results to better satisfy user information needs. However, existing methods show two notable limitations in improving user search satisfaction: inadequate modeling of multifaceted user intents and neglect of rich side information such as visual perception signals. To address these challenges, we propose the Rich-Media Re-Ranker framework, which aims to enhance user search satisfaction through multi-dimensional and fine-grained modeling. Our approach begins with a Query Planner that analyzes the sequence of query refinements within a session, decomposing the query into clear and complementary sub-queries to enable broader coverage of users' potential intents. Subsequently, moving beyond primary text content, we integrate richer side information of candidate results, including signals modeling visual content generated by the VLM-based evaluator. These comprehensive signals are then processed alongside carefully designed re-ranking principle that considers multiple facets, including content relevance and quality, information gain, information novelty, and the visual presentation of cover images. Then, the LLM-based re-ranker performs the holistic evaluation based on these principles and integrated signals. To enhance the scenario adaptability of the VLM-based evaluator and the LLM-based re-ranker, we further enhance their capabilities through multi-task reinforcement learning. The proposed framework has been deployed in a large-scale industrial search system, yielding substantial improvements in online user engagement rates and satisfaction metrics.

\end{abstract}




\maketitle

\section{Introduction}

In information search systems, the final results presented to users typically undergo three stages: retrieval, ranking, and re-ranking. The ranking stage commonly employs a pointwise scoring strategy~\cite{cheng2016wide, zhou2018deep, pi2020search, zhang2025onetrans, zhu2025rankmixer}, which focuses on evaluating individual results but fails to explicitly capture how each item affects the overall quality of the final list. For instance, this may lead to the presence of entries with highly homogeneous content positioned consecutively within the ranked list. To better satisfy user information needs and enhance their experience, the re-ranking stage aims to perform a more refined adjustment on candidate results after ranking, based on listwise awareness of the entire result set.

\begin{figure}[t]
    \centering

    \begin{tabular}{@{}c@{}}
        \includegraphics[width=0.47\textwidth]{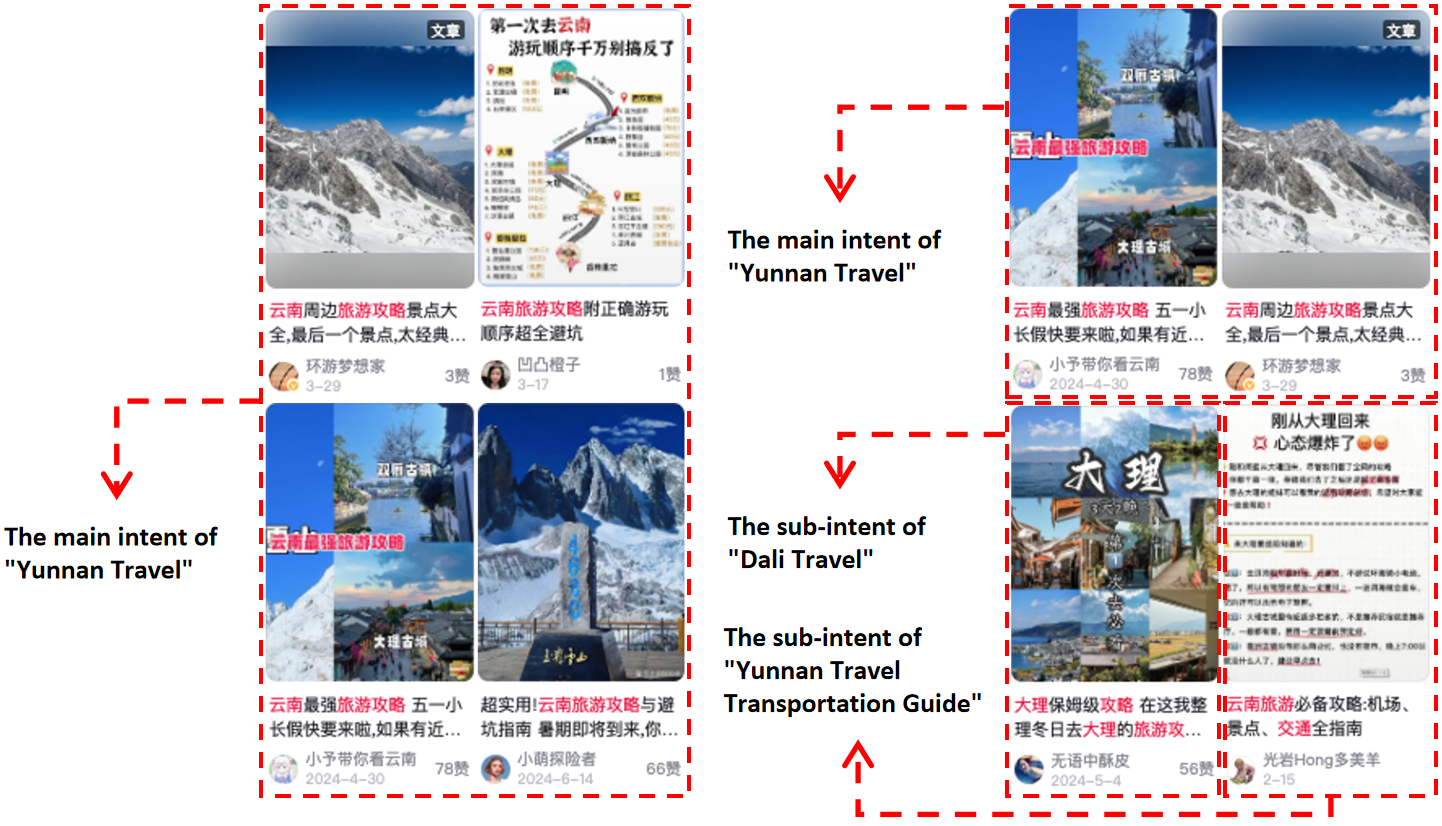} \\
        \parbox{0.47\textwidth}{\footnotesize \raggedright
            \textbf{(a)} \textbf{Query}: Yunnan Travel. \textbf{Left is original ranking:} Results primarily satisfy the main intent with severe content homogeneity. \textbf{Right is desired ranking:} Besides addressing the main intent, the diversified list covers sub-intents such as "Yunnan Dali Travel" and "Yunnan Travel Transportation Guide."
        }
    \end{tabular}

    \begin{tabular}{@{}c@{}}
        \includegraphics[width=0.47\textwidth]{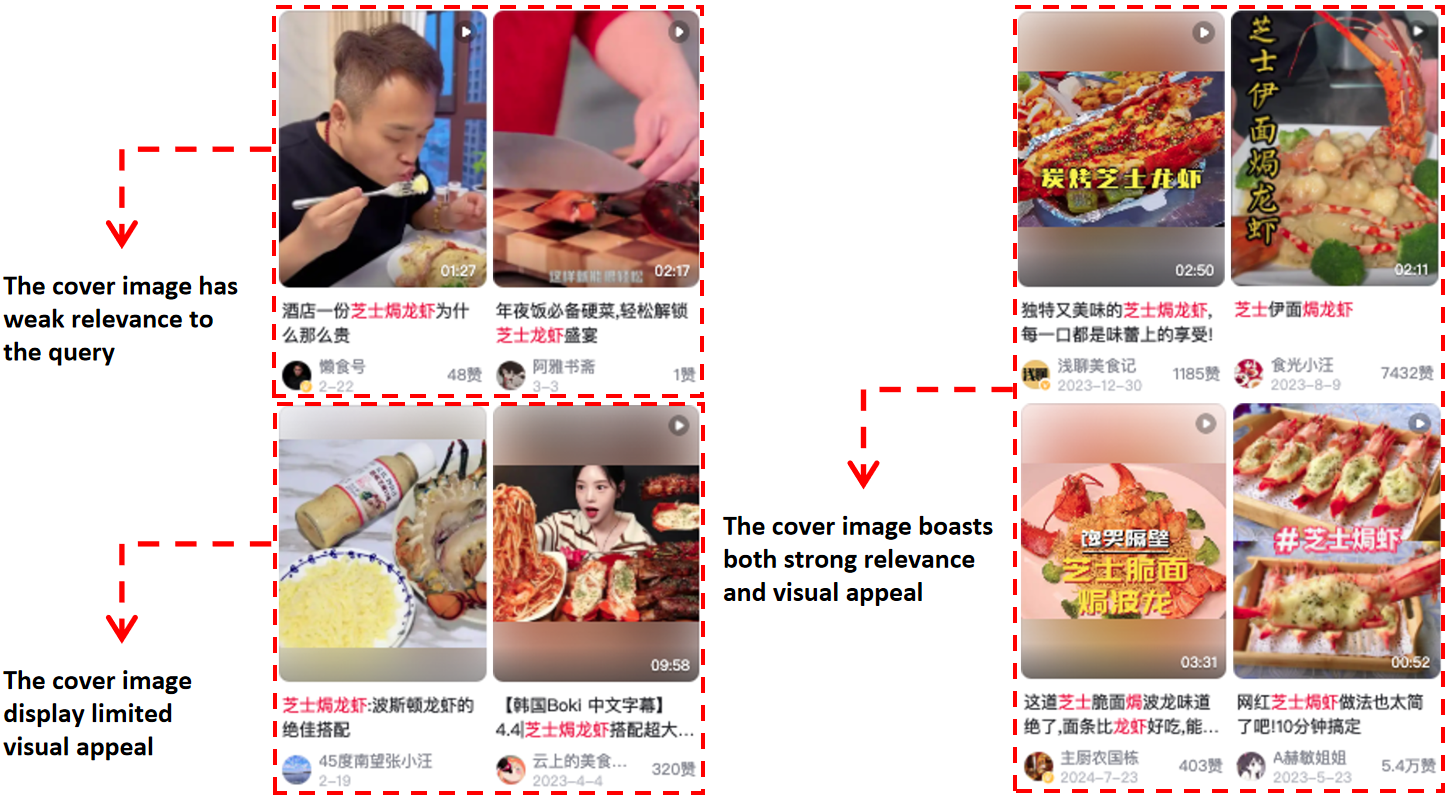} \\
        \parbox{0.47\textwidth}{\footnotesize \raggedright
            \textbf{(b)} \textbf{Query}: Cheese baked lobster. \textbf{Left is original ranking:} Cover images display limited visual appeal and weak relevance to the query. \textbf{Right is desired ranking:} All cover images exhibit stronger relevance and higher visual attractiveness.
        }
    \end{tabular}
    \captionsetup{skip=5pt}
    \caption{Comparison between original ranking and desired re-ranking results in rich-media search system.}
    \label{case_reranking}
\end{figure}

Traditional re-ranking approaches are commonly divided into two categories~\cite{lin2024discrete, ren2024non, wang2025nlgr}: generator-based methods and evaluator-based methods. Generator-based methods~\cite{liu2023generative,wang2025you,meng2025generative} re-score on items within the initial list by capturing how items may promote or suppress each other. Evaluator-based methods~\cite{ren2024non,lin2024discrete,wang2025nlgr} first produce multiple candidate lists from the large permutation space, and measure these candidates with a learned utility function. Recent advances in Large Language Models (LLMs) have inspired a new line of LLM-based re-ranking methods~\cite{liu2025coranking,jin2025rankflow,zhuang2024setwise,gao2025llm4rerank,zhang2025rearank,zhuang2025rank,liu2025reasonrank}, which utilize strong understanding and generation capabilities to directly reason over candidates and output the final ranked list, showing encouraging performance.

Despite demonstrating certain effectiveness, they face two significant challenges in modern search scenarios that remain to be solved, especially amid the trend toward rich-media content: \textbf{(1) User queries often encapsulate multidimensional latent intents.} As illustrated in Figure~\ref{case_reranking}(a), taking the query “Yunnan Travel” as an example, besides the primary tourism intent, the underlying needs also include more specific sub-intents such as “Yunnan Dali Travel” and “Yunnan Travel Transportation Guide.” Existing methods treat the raw query as the sole reference for intent understanding, showing the limitations in exploring underlying needs, and still lack an appropriate unified optimization objective to holistically address users’ complex and multi-aspect information needs. \textbf{(2) Rich side information associated with candidate results, particularly the appeal of cover images, critically influences users' first impressions and click decisions}, illustrated in Figure~\ref{case_reranking}(b). Most existing methods primarily focus on textual information, which restricts their capability in modern search scenarios. Neglecting deep parsing and satisfying user intents and modeling rich side information hinders existing re-ranking technologies from adapting to increasingly complex information-seeking behaviors and multimodal interaction preferences.

To address these challenges, we propose the Rich-Media Re-Ranker framework, performing fine-grained re-ranking in search systems by enhancing user intent modeling and leveraging rich side information. Our approach begins with a session-aware Query Planner, which leverages the user session context to analyze sequences of query refinements and decomposes complex and broad-needs queries into clear and complementary sub-queries, enriching the candidate pool to enable broader coverage of users' potential intents. Subsequently, moving beyond primary text content, we integrate richer fine-grained side information of candidate results, which includes signals modeling visual content generated by a VLM-based evaluator. These comprehensive signals are then processed by an LLM-based re-ranker to perform a holistic evaluation based on carefully designed re-ranking principle, which considers multiple facets, including content relevance and quality, information gain, information novelty, and the visual presentation of cover images. To enhance the scenario adaptability of both the VLM-based evaluator and the LLM-based re-ranker, we further boost their capabilities through multi-task reinforcement learning, designing reward functions for specific task objectives. Our contributions are as follows:
\begin{itemize}[leftmargin=0.3cm]
    \item We design a session-aware Query Planner that deconstructs complex and broad-needs user queries into executable, complementary sub-intentions by analyzing query refinement sequences within a session, facilitating fine-grained and comprehensive modeling of multifaceted user intents.
    \item We introduce rich side information integration, and utilize a post-trained VLM via reinforcement learning to deliver a multidimensional evaluation of relevance and quality, which is then utilized as a critical feature in the subsequent re-ranking process.
    \item We construct an LLM-based re-ranker, which is further enhanced through reinforcement learning to make final decisions by synthesizing content relevance and quality, overall information gain, information novelty, and visual presentation of cover images.
    \item Extensive experiments show that our framework achieves substantial improvements over state-of-the-art baselines. Moreover, it has been successfully deployed in a large-scale industrial search system, where it yields significant enhancements in online user satisfaction and engagement rates.
\end{itemize}

\section{Related Works}

\subsection{Traditional Re-Ranking Methods}

Previous research on re-ranking can be classified into two categories: generator-based methods~\cite{pei2019personalized,xi2022multi,bello2018seq2slate,ai2018learning,zhuang2018globally,pang2020setrank,liu2023generative,wang2025you,meng2025generative} and evaluator-based methods~\cite{feng2021revisit,shi2023pier,ren2024non,lin2024discrete,feng2021grn,xi2021context,wang2025nlgr,yang2025comprehensive,jiang2018beyond,liu2021variation,zhang2025generation,mao2025robust}. The former operates as a single-stage paradigm, whose core idea is to leverage context-aware models like Transformers to perform interactive modeling and re-scoring of items within the initial list. By learning how items mutually promote or suppress each other, these methods refine the initial scores of individual items and subsequently generate the sequence via greedy sorting. The latter employs a two-stage generator-evaluator framework, the generator is tasked with producing plausible candidate lists from the exponentially large permutation space, while the evaluator comprehensively assesses these candidates against a learned list utility function to ultimately select the best-performing sequence. Despite their differences, they share common limitations. First, they often fail to adequately model multifaceted user intent. Many approaches either overlook the semantic richness present in user queries by relying exclusively on single-content-side ranking or employ relatively simple query modeling approaches. This limits their capacity to comprehensively understand and fulfill users’ underlying multi-aspect information needs, often resulting in re-ranked outputs that improve merely in superficial diversity rather than true alignment with user expectations. Second, their decision-making process typically lacks transparency. Operating mainly through implicit prediction mechanisms, these methods provide no interpretable reasoning path, making it difficult to assess whether and how substantially user needs have been satisfied.

\subsection{Large Language Models for Re-Ranking}

In recent years, LLM-based re-ranking techniques have become increasingly popular~\cite{abdallah2025dear,wang2025realm,liu2025coranking,jin2025rankflow,zhuang2024setwise,gao2025llm4rerank,zhang2025rearank,zhuang2025rank,liu2025reasonrank}. These methods leverage the powerful instruction-following and generative capabilities of LLMs to perform global evaluation of entire candidate document lists and generate final ranked results. For instance, LLMReranker~\cite{zhuang2024setwise}, LLM4Rerank~\cite{gao2025llm4rerank} and RankFlow~\cite{jin2025rankflow} utilize the zero-shot ability of LLMs to conduct list-wise re-ranking of candidate documents; while ReaRank~\cite{zhang2025rearank}, Rank-R1~\cite{zhuang2025rank} and ReasonRank~\cite{liu2025reasonrank} employ supervised fine-tuning and reinforcement learning to specifically enhance their document re-ranking abilities. However, these approaches fail to adequately address core challenges emerging in modern search environments. First, user queries tend to be increasingly concise but with multiple intents, where a single request often implies multi-layered composite needs. Second, the importance of visual characteristics in search result entries is becoming increasingly prominent, especially in interfaces dominated by rich media streams or card-style layouts, where the attractiveness of cover images has become a critical factor influencing user decisions. Unfortunately, mainstream LLM-based re-ranking technologies remain largely confined to text semantic relevance matching~\cite{zhang2025rearank,zhuang2025rank,liu2025reasonrank}, lacking systematic decomposition of users' deep and diverse intents while failing to incorporate modeling of result appeal into the decision-making process. Although MM-R5~\cite{xu2025mm} and GMR~\cite{dai2025supervised} employ VLMs for cross-modal re-ranking, their applicability is limited to single-modality candidate content, making them unsuitable for the composite-modal content in the search system. This limitation hinders existing methods from effectively adapting to users' increasingly complex information needs and multimodal interaction preferences, thereby limiting their practical utility and user experience enhancement in real-world scenarios.

\section{Methodology}

We propose Rich-Media Re-Ranker, addressing limitations in existing methods through fine-grained intent modeling and rich side information integration. Our methodology begins with a session-aware Query Planner that decomposes queries into clear and complementary sub-queries, expanding coverage of potential intents (Section~\ref{sec:query_planner}). Then, the VLM-based evaluator assesses cover image relevance and quality, which are combined with textual content and behavioral features, to feed into the LLM-based re-ranker. The re-ranker holistically evaluates candidates across content relevance, quality, information novelty, and visual presentation to produce interpretable rankings (Section~\ref{sec:re-ranking}). Finally, multi-task reinforcement learning for VLM and LLM enhances both the visual assessment and re-ranking capabilities in complex search scenarios (Section~\ref{sec:training}). 

\begin{figure}[t]
\centering
\includegraphics[width=1.0\linewidth]{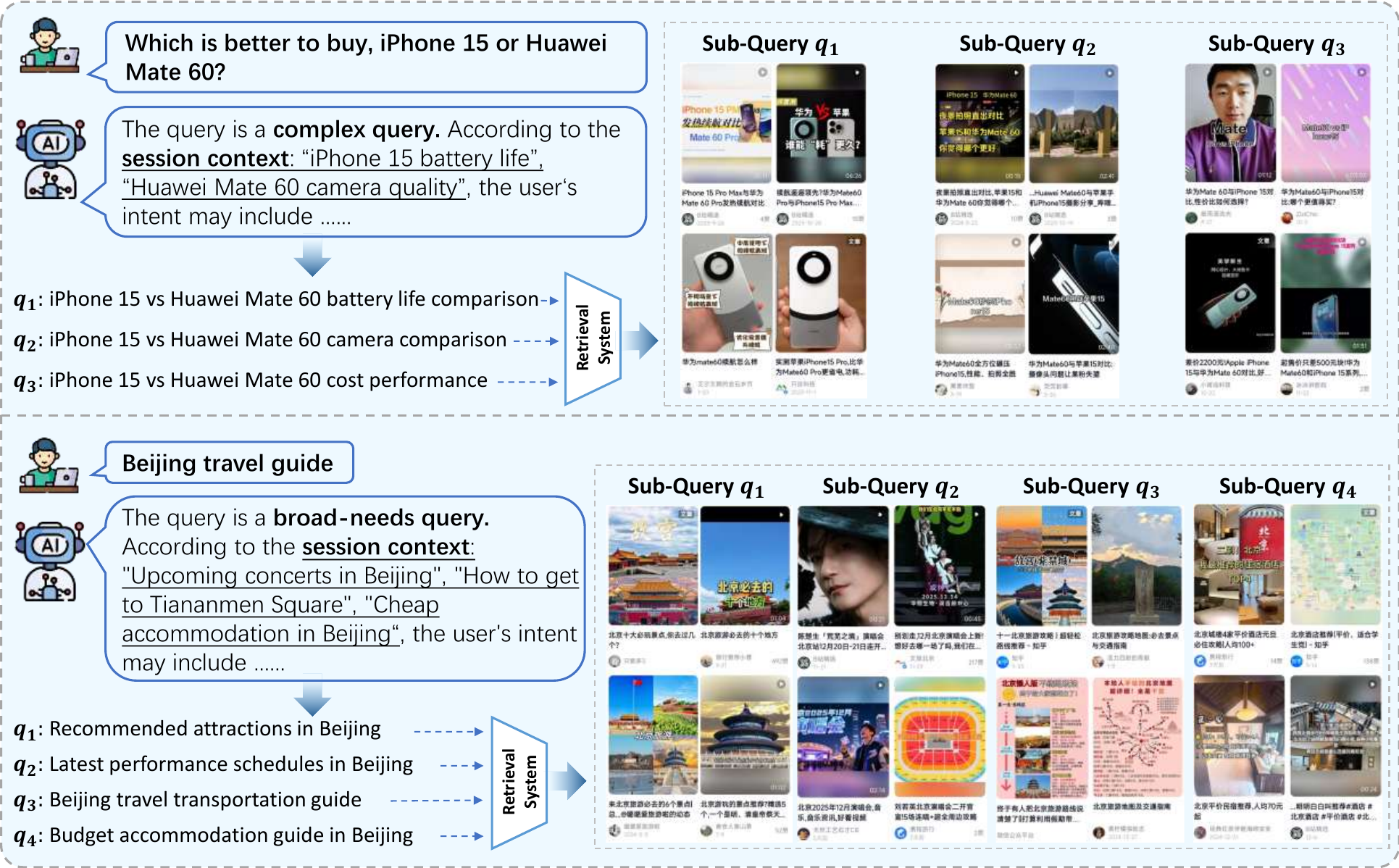} 
\captionsetup{skip=5pt}
\caption{The workflow of Query Planner.}
\label{query_planner}
\end{figure}

\subsection{Session-aware Query Planner}
\label{sec:query_planner}
The Query Planner is designed to address the limitation of inadequate modeling of multifaceted user intent in existing methods. Given an original user query $Q$ and its session context $S = \{hist_{1}, hist_{2}, \dots\}$ representing the user's query history within the current search session, the Query Planner conducts a deep semantic analysis via LLM. Its core objective is to leverage session context $S$ to infer genuine user intent, then systematically deconstruct ambiguous or compound information needs into a coordinated set of simpler, executable sub-queries, and obtain a candidate set of results that can meet the diverse needs of users.

Firstly, the Query Planner analyzes the query and classifies it into one of three predefined types: complex query, broad-needs query, or simple query. For complex and broad-needs queries, the planner leverages session context $S$ to infer genuine user intent. Based on the query type and inferred session intent, the planner generates a set of sub-queries $P(Q) = \{q_1(\mathbf{d}_1), q_2(\mathbf{d}_2), \dots\}$ where each $q_i$ targets a specific facet of the original information need, and $\mathbf{d}_i$ denotes intent dimensions including \textit{High Freshness}, \textit{Authoritativeness}, and \textit{Personal Experience}. As shown in Figure~\ref{query_planner}, the process rule as follow:
\begin{itemize}[leftmargin=0.3cm]
    \item \textbf{Complex Query}: Identify portions requiring comparisons or multi-step information gathering, breaking them down into independently searchable sub-queries inferred from session context. For example, the query "Which is better to buy, iPhone 15 or Huawei Mate 60?" is classified as a Complex Query. Given session context $S=\{$"iPhone 15 battery life", "Huawei Mate 60 camera quality"$\}$, it is decomposed into sub-queries: "iPhone 15 vs Huawei Mate 60 battery life comparison" (Authoritativeness, Personal Experience), "iPhone 15 vs Huawei Mate 60 camera comparison" (Authoritativeness, Personal Experience), and "iPhone 15 vs Huawei Mate 60 cost performance" (Personal Experience).
    \item \textbf{Broad-needs Query}: Recognize and extend possible multidimensional aspects based on session context, generating sub-queries that broadly yet non-redundantly cover potential intents. For example, the query "Beijing travel guide" is classified as a Broad-needs Query. Given session context $S=\{$"Upcoming concerts in Beijing", "How to get to Tiananmen Square", "Cheap accommodation in Beijing"$\}$, it is expanded into sub-queries: "Recommended attractions in Beijing" (Personal Experience), "Latest performance schedules in Beijing" (Authoritativeness, High Freshness), "Beijing travel transportation guide" (Authoritativeness), and "Budget accommodation guide in Beijing" (Personal Experience).
    \item \textbf{Simple Query}: For explicit and directly answerable queries, rewrite it appropriately if obvious typo error is detected; otherwise, preserve the original query. Session context is not considered for this query type. For example, the query "Python tutoral" is classified as a simple query and rewritten as "Python tutorial" (Authoritativeness), while the query "weather today" is classified as a simple query and remains unchanged as "weather today" (Authoritativeness, High Freshness). 
\end{itemize}

Each sub-query $q \in P(Q)$ is then independently sent to the retrieval system, which returns a list of top-$k$ candidate documents, denoted as $R(q, k)$. All retrieved results are merged to form the final candidate set $C(Q)$ for the subsequent re-ranking stage: $C(Q) = \bigcup_{q \in Q} R(q, k) = \{doc_1, doc_2, \dots \}$. The intent dimensions $\mathbf{d}$ associated with each sub-query are passed to the re-ranking stage, where they serve as explicit signals to prioritize certain characteristics in the results. For example, $\textit{High Freshness}$ indicates prioritizing recent content, $\textit{Authoritativeness}$ emphasizes authoritative sources, and $\textit{Personal Experience}$ highlights user testimonials.

By decomposing the original query in this manner, the Query Planner ensures that the input to the re-ranker comprehensively covers the diverse aspects of the user's potential intents, laying a solid foundation for high-quality, intent-aligned result re-ranking.

\subsection{Re-ranking with Integrated Signal}
\label{sec:re-ranking}
To overcome limitations of content text-only re-ranking approaches, we establish a comprehensive framework that integrates rich side information and multi-modal signals under unified re-ranking principles, which contains three critical components: augmented side information beyond the textual content, visual value assessment of cover images, and the multifaceted re-ranking principle, enabling holistic candidate evaluation.

\subsubsection{Augmented Side Information}
\label{sec:side_info}
Beyond primary textual content such as title and content, we integrate richer contextual signals that significantly influence user decision-making. Specifically, publication time is incorporated to prioritize time-sensitive queries, particularly in domains like news or trending topics. Then, we include authentic user posterior feedback comprising click-through rates and content completion rates, which captures implicit user preferences and content utility. These behavioral signals establish a direct connection to genuine user engagement patterns, enabling the model to transcend purely prior assessments.

\subsubsection{Cover Image Value Assessment}

To overcome the limitation of existing re-ranking methods that overlook visual characteristics, we argue that assessing cover images is critical because user interactions with search results are profoundly influenced by the cover image they see first. 

\begin{figure}[t]
    \centering
    \begin{minipage}{1.0\linewidth}
        \centering
        \includegraphics[width=\textwidth]{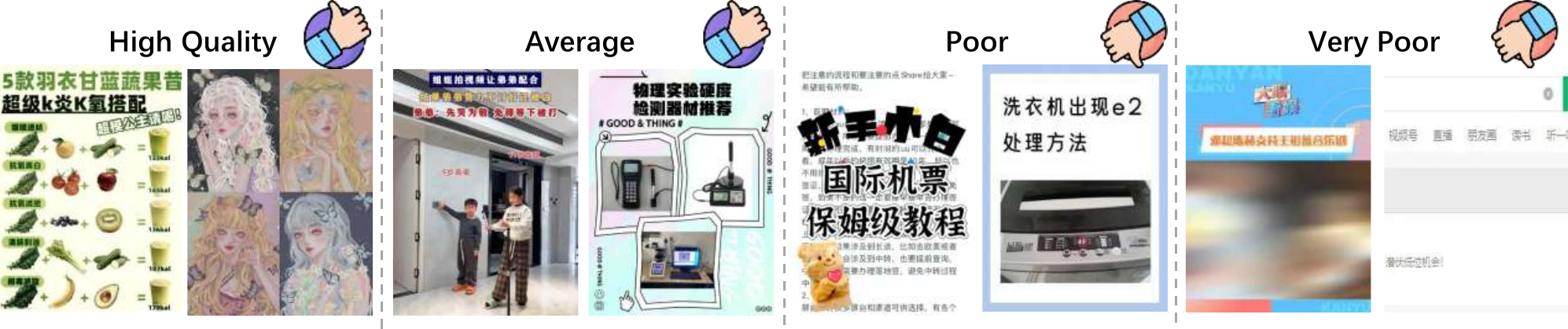}
        \\ 
        \small (a) Quality dimension.
    \end{minipage}
    \bigskip 
    \begin{minipage}{1.0\linewidth}
        \centering
        \includegraphics[width=\textwidth]{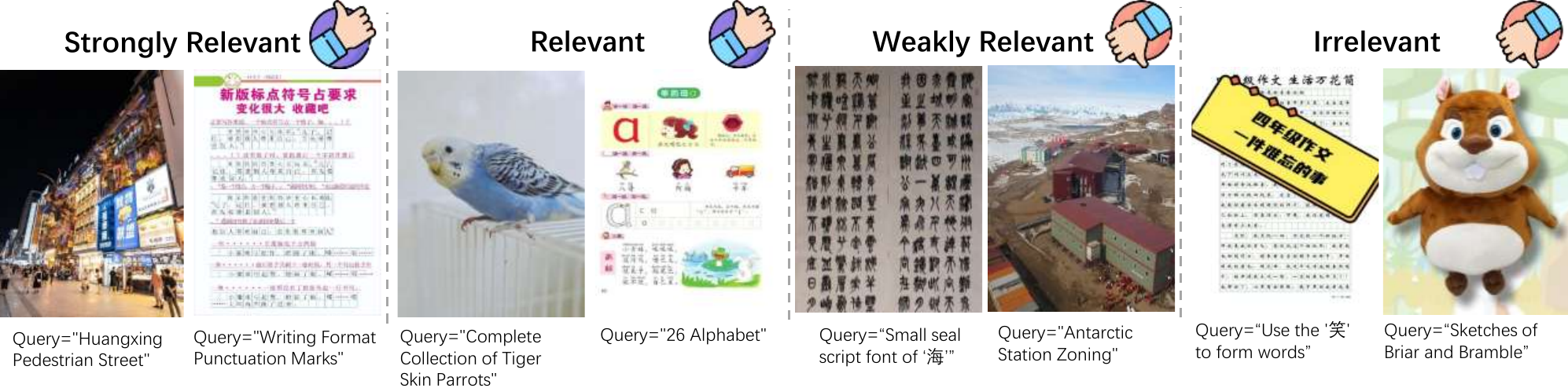}
        \\
        \small (b) Relevance dimension.
    \end{minipage}
    \captionsetup{skip=0pt}
    \caption{Example of cover image value.}
    \label{cover_value}
\end{figure}

When browsing results, users' underlying behavioral patterns can be viewed through two primary pathways: those actively looking for concrete answers and those engaging in exploratory discovery. For the former, a cover image containing direct answers or key information holds exceptional value and high click-through potential, which we term as relevance. For the latter, the user's click decision is strongly affected by the immediate visual appeal of the cover, including attributes such as aesthetic composition, color vibrancy, layout harmony, and the absence of disruptive elements like watermarks or blurring, which we collectively term as quality. To conduct quantitative evaluation specifically along these two decisive dimensions, as shown in Figure~\ref{cover_value}, we formalize both relevance and quality into four tier grading scales: relevance is rated as Strongly Relevant, Relevant, Weakly Relevant, or Irrelevant, corresponding to scores of 4, 3, 2, and 1; similarly, quality is defined as High Quality, Average, Poor, or Very Poor, mapped to the same numeric scale. We utilizes VLM as evaluator, for each candidate $doc_i \in C(Q)$ of query $Q$:
\begin{align}
    \text{S}_{\text{rel}}(c_i) &= \text{VLM}(\mathcal{P}_{\text{rel}}; Q, Img_i), \label{eq:rel} \\
    \text{S}_{\text{qual}}(c_i) &= \text{VLM}(\mathcal{P}_{\text{qual}}; Img_i). \label{eq:qual}
\end{align}
where $Img_i$ denotes the cover image of $doc_i$, $\mathcal{P}_{\text{rel}}$ constitutes the rule set for relevance assessment defining the semantic alignment criteria between query and cover image, and $\mathcal{P}_{\text{qual}}$ represents the quality assessment rule set specifying comprehensive visual evaluation dimensions including content value, image completeness, clarity, aesthetic appeal, layout comfort, color harmony, and element richness, providing in the Appendix~\ref{sec:prompt_vlm}. These visual signals are subsequently integrated into the final re-ranking process.

\begin{figure*}[t]
\centering
\includegraphics[width=1.0\textwidth]{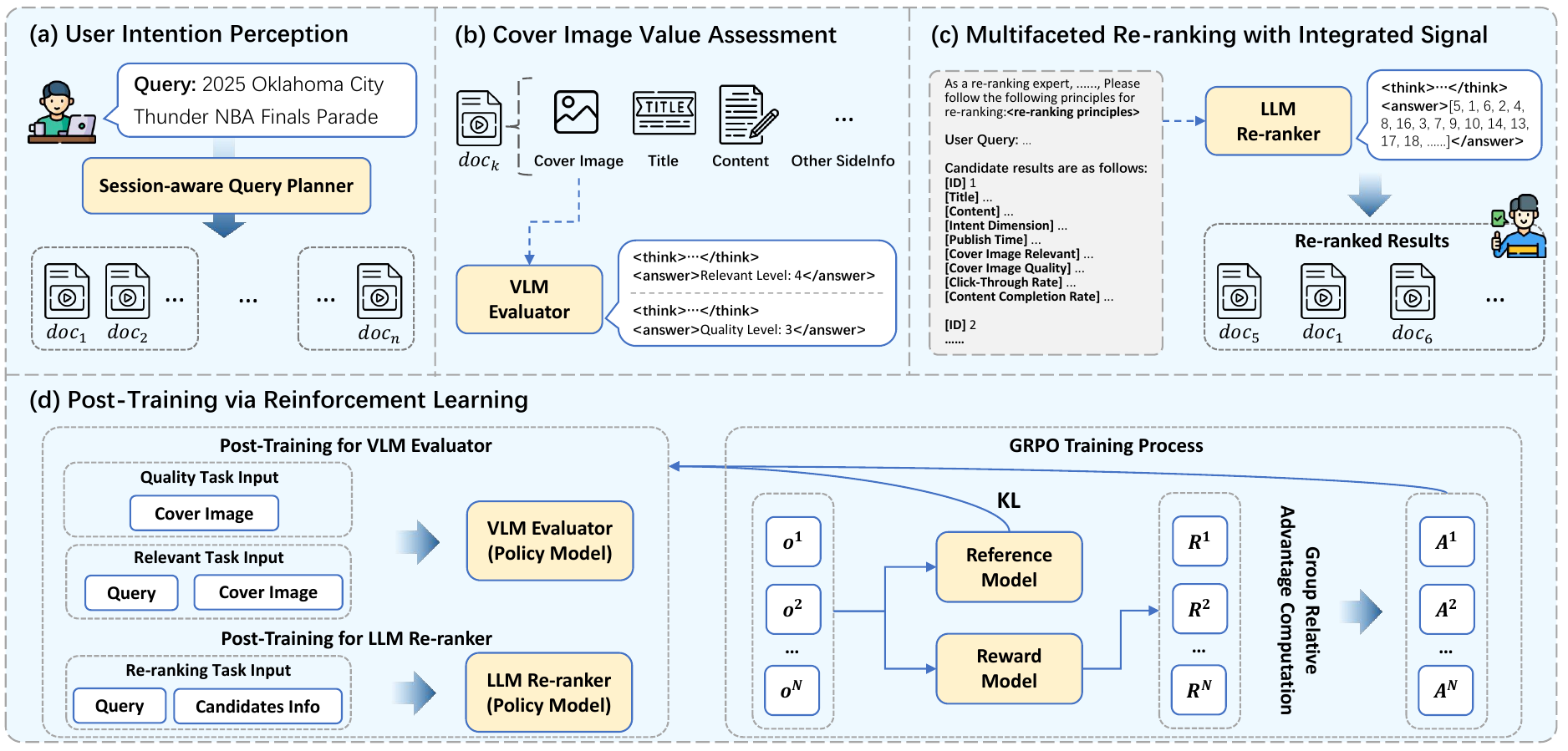} 
\captionsetup{skip=5pt}
\caption{Framework of our Rich-Media Re-Ranker. (a) We first leverage a Session-aware Query Planner to capture the user's multi-dimensional intent. The top-k results from decomposed sub-queries are combined as the candidate set for re-ranking, which improves intent coverage. (b) The VLM Evaluator assesses the relevance and quality of document cover images, integrates this with side information, providing signals for subsequent re-ranking. (c) All documents are organized in a structured manner according to the designed re-ranking principle. Considering multiple dimensions, the LLM Re-ranker produces the document rank that aligns with user satisfaction. (d) We enhance the capabilities of the VLM Evaluator and LLM Re-ranker in real-world data and tasks through multi-task reinforcement learning.}
\end{figure*}

\subsubsection{Multifaceted Re-ranking Principle}

Our re-ranking strategy integrates multiple criteria, including content relevance and quality, information gain, information novelty, and visual presentation of cover images. Specifically, we instruct the LLM to give high priority to candidates that match the main intent of the user query. Then, we emphasize diversity in content to maximize informative gains while reducing redundancy. When the information utility among items appears comparable, the model incorporates the cover image relevance and quality scores estimated by the VLM evaluator as important signals. Furthermore, for clearly time-sensitive queries, such as recent news, content recency is prioritized accordingly. Importantly, the LLM's strong reasoning ability enables it to articulate the rationale behind each ranking decision, contributing interpretability to the entire re-ranking process. Formally, the LLM re-ranker operates in a listwise manner over the candidate set $C(Q)$, with each candidate $doc_i$ represented as:
\begin{align}
    \label{eq:reranker_input}
    \Psi_i = \left( \text{Title}_i, \text{Content}_i, \text{SideInfo}_i, d_i, \text{S}_{\text{rel}}(c_i), \text{S}_{\text{qual}}(c_i) \right),
\end{align}
where $\text{SideInfo}_i$ specifically includes the information introduced in Section~\ref{sec:side_info}, $d_i$ is the intent dimension of the corresponding sub-query, as described in Section~\ref{sec:query_planner}. The re-ranking orders are computed as:
\begin{align}
    \label{eq:reranker_output}
    \left[ s_1, s_2, \ldots, s_n \right] = \text{LLM}\left( \mathcal{P}_{\text{re-ranking}}; Q, \left\{ \Psi_k \right\}_{k=1}^n \right),
\end{align}
where $\mathcal{P}_{\text{re-ranking}}$ represents the designed re-ranking principle, providing in the Appendix~\ref{sec:prompt_rerank}.

\subsection{Post-Training via Reinforcement Learning}
\label{sec:training}
While general-purpose VLMs demonstrate impressive zero-shot capabilities across diverse tasks, we observe significantly degraded performance when directly applying them to assess cover image relevance and quality in our service scenario. Similarly, LLMs exhibit comparable limitations in zero-shot re-ranking tasks. To address these challenges, we employ specialized enhancement for both components via multi-task reinforcement learning, ensuring they develop robust task-specific competencies.

\subsubsection{High Quality Data Preparation}

For the cover image evaluation task, we curate a dataset of human-annotated assessments across both relevance and quality dimensions, with approximately 500 images per tier for VLM training, providing a reliable graded benchmark. For the re-ranking task, we leverage DeepSeek-R1~\cite{guo2025deepseek} as the teacher model to synthesize high-quality ranked lists. To ensure result reliability, for each query and its associated candidate set, we repeatedly randomize the input order of candidates and generate responses, then compute Rank-Biased Overlap (RBO) across all results with the same query. We retain only those results exhibiting extremely high internal consistency. Additionally, to avoid excessive verbosity in the chain of thought, which may impede stable learning, we compress the overlong chain of thought using DeepSeek-V3~\cite{liu2024deepseek}, retaining core logical elaborations.

\subsubsection{Post-Training via GRPO}

Group Relative Policy Optimization (GRPO)~\cite{shao2024deepseekmath} provides a unified framework for optimizing cover image assessment and re-ranking tasks through relative reward mechanisms. Specifically, the policy model $\pi_\theta$ generates a group of $N$ responses $\{o^{1}, o^{2}, \ldots, o^{N}\}$ per input instance, each accompanied by explicit reasoning steps. Each response is then evaluated using its corresponding reward functions and computing rewards for each group of responses. After computing relative advantage scores within each response group, the policy model is optimized jointly via the multi-task GRPO:

\begin{equation}
    \begin{split}
        & \mathcal{L}_{\text{GRPO}} = \sum_{t \in \mathcal{T}} \Bigg[ - \frac{1}{N} \sum_{i=1}^{N} \bigg( \min\left( \frac{\pi_\theta(o_{t}^{i}|x_t)}{\pi_{\text{old}}(o_{t}^{i}|x_t)} \cdot \hat{A}_{t}^{i}, \right. \\
        & \left. \text{clip}\left( \frac{\pi_\theta(o_{t}^{i}|x_t)}{\pi_{\text{old}}(o_{t}^{i}|x_t)}, 1-\epsilon, 1+\epsilon \right) \cdot \hat{A}_{t}^{i} \right) - \beta \cdot \text{D}_{\text{KL}}\left( \pi_\theta \parallel \pi_{\text{old}}\right) \bigg) \Bigg],
    \end{split}
    \label{eq:grpo_loss}
\end{equation}

\begin{equation}
    A_{t}^{i} = \frac{R_t(o_{t}^{i}) - \text{mean}(\{R_t(o_{t}^{j})\}_{j=1}^N)}{\text{std}(\{R_t(o_{t}^{j})\}_{j=1}^N)},
\end{equation}
where $\mathcal{T}$ denotes the task type $\{\text{rel}, \text{qual}\} \setminus \{\text{re-rank}\}$, $x_t$ represents the input for task $t$, $o_{t}^{i}$ is the $i$-th response in the group of $N$ outputs generated by policy model $\pi_\theta$, with $\pi_\theta$ and $\pi_{\text{old}}$ being the current and reference policies respectively, $R_t(o_{t}^{i})$ indicates the task-specific reward, $\epsilon$ serves as the clipping range hyperparameter preventing unstable policy updates, and $\beta$ acts as the KL regularization coefficient constraining policy drift.

The VLM evaluator addresses two distinct assessment tasks with differentiated input requirements. For cover relevance assessment, inputs incorporate the cover image combined with the user query and relevance evaluation criteria, as in ~Eq.\eqref{eq:rel}. For cover quality assessment, inputs consist solely of the cover image and quality evaluation guidelines, as in~Eq.\eqref{eq:qual}. Both tasks produce identical output structures, containing explicit reasoning chains and integer ratings conforming to predefined four-tier grading scales.

The re-ranking task processes inputs comprising user queries paired with candidate content sets, where each candidate includes textual content, side infomation, and VLM-generated cover image assessment scores, as in~Eq.\eqref{eq:reranker_input}. Prior to reinforcement learning, given that vanilla LLM lack profound training in re-ranking rules, we execute a brief cold-start supervised fine-tuning (SFT) on the synthesized dataset, allowing the re-ranker to rapidly acquire the target task paradigm, output style, and formatting norms, thus providing a foundation for steadier performance gains and reducing early-phase inefficiencies in reinforcement learning.

\subsubsection{Reward Design for VLM Evaluator}

Appropriate reward design is crucial for guiding the model to learn task patterns and effective reasoning. We carefully design the following reward functions to enhance the relevance and quality evaluation performance of VLM on content cover images.

\begin{itemize}[leftmargin=0.3cm]
    \item \textbf{Format Reward} $r^{\text{format}}_\text{rel/qual}$: This reward evaluates whether the chain of thought and final answer can be correctly extracted from the generated response. First, checking whether the response contains paired <think></think> and <answer></answer> tags. Then, verifying whether the answer inside <answer></answer> tag can be correctly extracted and whether it conforms to the specified integer format (corresponding to the rating levels):
    \begin{equation}
        r^{\text{format}}_\text{rel/qual} =
            \begin{cases}
            0.0, & \text{No valid answer} \\
            0.3, & \text{Non-integer answer} \\
            0.5, & \text{Valid answer}.
            \end{cases}
    \end{equation} 
    
    \item \textbf{Task Reward} $r^{\text{task}}_\text{rel/qual}$: For each assessment task (relevance or quality), the model is required to predict the grade category. Let $\text{grade}_{\text{pred}}$ denote the predicted grade for the $i$-th response, and $\text{grade}_{\text{gt}}$ represent the ground-truth grade. The score reward $r_{\text{scr}}$ is defined based on the absolute difference between predicted and ground-truth numerical scores:     
    \begin{equation}
        r^{\text{task}}_\text{rel/qual} =
            \begin{cases}
                1.0, & \text{if } |\text{grade}_{\text{pred}} - \text{grade}_{\text{gt}}| = 0, \\
                0.7, & \text{if } |\text{grade}_{\text{pred}} - \text{grade}_{\text{gt}}| = 1, \text{no cross boundary} \\
                0.4, & \text{if } |\text{grade}_{\text{pred}} - \text{grade}_{\text{gt}}| = 1, \text{cross boundary} \\
                0, & \text{if } |\text{grade}_{\text{pred}} - \text{grade}_{\text{gt}}| \geq 2.
            \end{cases}
    \end{equation} 
\end{itemize}

Since the four-level scoring scale inherently forms two natural preference intervals, with grades 1\textasciitilde2 indicating unfavorable evaluations and grades 3\textasciitilde4 reflecting favorable ones, we impose stronger penalties when predictions incorrectly cross these meaningful boundaries. For instance, predicting 3 when the true label is 2 incurs a greater loss despite the minimal numerical difference. In contrast, one-level deviations that remain within the same preference interval are treated more tolerantly.

\subsubsection{Reward Design for LLM Re-ranker}
The reward function designed for the LLM re-ranker aims to enhance its capability for complex re-ranking of candidate results, with a emphasis on practical application priorities (e.g., Top-4 results).

\begin{itemize}[leftmargin=0.3cm]
    \item \textbf{Format Reward} $r^{\text{format}}_\text{re-rank}$: The fundamental format verification follows the same principles as before, with the key distinction lying in ensuring the usability of the generated answers as valid re-ranking results. Specifically, penalties are imposed when: (1) the final answer cannot be converted into a List, or (2) the generated re-ranking list contains duplicate IDs, invalid IDs, or missing IDs, making the output unusable:
    \begin{equation}
        r^{\text{format}}_\text{re-rank} =
            \begin{cases}
            0.0, & \text{No valid answer} \\
            0.2, & \text{Some IDs duplicate, invalid, or missing} \\
            0.5, & \text{Valid answer}.
            \end{cases}
    \end{equation}      
    
    \item \textbf{Task Reward} $r^{\text{task}}_{\text{re-rank}}$: For the output re-ranked list, inspired by ~\cite{liu2025reasonrank}, we comprehensively consider NDCG@K, Recall@K, and RBO. However, since in search systems, top-ranked results often determine the overall user experience, in addition to considering the entire list, we additionally incorporate enhanced supervision specifically for top-ranked results:     
    \begin{align}
        r^{\text{task}}_{\text{re-rank}} = &\ \gamma_1 \cdot \mathrm{NDCG}@10 + \gamma_2 \cdot \mathrm{Recall}@10 + \gamma_3 \cdot \mathrm{RBO} \\
        & + \gamma_4 \cdot \mathrm{NDCG}@4 + \gamma_5 \cdot \mathrm{Recall}@4
    \end{align} 
    where $\gamma_1$ through $\gamma_5$ are trade-off parameters serving to both balance the relative contribution of each reward component and normalize the aggregate reward for stabilized training.
\end{itemize}

For responses of each input data, the overall reward for the $i$-th response is computed as $R_t^i =r_t^\text{format}(o^{i}) + r_t^\text{task}(o^{i})$, where $t$ is the task type. This joint training strategy enables the model to facilitate robust performance in corresponding assessment tasks.


\section{Offline Evaluation}

\setlength{\equalcolwidth}{0.0375\textwidth} 

\begin{table*}[t]
\renewcommand{\arraystretch}{1.1}
  \centering
  \captionsetup{skip=5pt}
  \caption{The performance (\%) comparison. The best results are indicated in \textbf{bold} and the runner-ups are \underline{underlined}.
}  
    \resizebox{\linewidth}{!}{
    \setlength{\tabcolsep}{2.1pt}
    \begin{tabular}{@{}cc*{21}{>{\centering\arraybackslash}p{\equalcolwidth}}@{}}
    \toprule
    \multirow{2}{*}{\textbf{Method}} 
     & \multicolumn{5}{c}{\textbf{All Query}}
     & \multicolumn{5}{c}{\textbf{Complex Query}} 
     & \multicolumn{5}{c}{\textbf{Broad-needs Query}} 
     & \multicolumn{5}{c}{\textbf{Simple Query}} \\
    \cmidrule(lr){2-6} \cmidrule(lr){7-11} \cmidrule(lr){12-16} \cmidrule(lr){17-21}
     & {N@4} & {N@10} & {R@4} & {R@10} & {RBO} 
     & {N@4} & {N@10} & {R@4} & {R@10} & {RBO} 
     & {N@4} & {N@10} & {R@4} & {R@10} & {RBO} 
     & {N@4} & {N@10} & {R@4} & {R@10} & {RBO} \\
    \midrule

    \textbf{Rank-R1} 
    & \multicolumn{1}{c}{82.4} 
    & \multicolumn{1}{c}{74.3} 
    & \multicolumn{1}{c}{62.9} 
    & \multicolumn{1}{c}{55.6} 
    & \multicolumn{1}{c}{58.9}
    & \multicolumn{1}{c}{81.1} 
    & \multicolumn{1}{c}{74.1} 
    & \multicolumn{1}{c}{59.9} 
    & \multicolumn{1}{c}{56.3} 
    & \multicolumn{1}{c}{58.4} 
    & \multicolumn{1}{c}{77.9} 
    & \multicolumn{1}{c}{72.9} 
    & \multicolumn{1}{c}{54.3} 
    & \multicolumn{1}{c}{55.2} 
    & \multicolumn{1}{c}{56.4} 
    & \multicolumn{1}{c}{90.8} 
    & \multicolumn{1}{c}{76.1} 
    & \multicolumn{1}{c}{79.3} 
    & \multicolumn{1}{c}{54.9} 
    & \multicolumn{1}{c}{62.8}  \\

 \hdashline

  \makecell[c]{+ Re-ranking Principles}
    & \multicolumn{1}{c}{85.5} 
    & \multicolumn{1}{c}{76.8} 
    & \multicolumn{1}{c}{66.0} 
    & \multicolumn{1}{c}{58.5} 
    & \multicolumn{1}{c}{61.4}
    & \multicolumn{1}{c}{87.1} 
    & \multicolumn{1}{c}{78.5} 
    & \multicolumn{1}{c}{65.8} 
    & \multicolumn{1}{c}{61.0} 
    & \multicolumn{1}{c}{63.2} 
    & \multicolumn{1}{c}{81.1} 
    & \multicolumn{1}{c}{76.2} 
    & \multicolumn{1}{c}{57.5} 
    & \multicolumn{1}{c}{59.3} 
    & \multicolumn{1}{c}{59.3} 
    & \multicolumn{1}{c}{91.1} 
    & \multicolumn{1}{c}{77.0} 
    & \multicolumn{1}{c}{79.7} 
    & \multicolumn{1}{c}{55.7} 
    & \multicolumn{1}{c}{63.3}  \\

 \hdashline

  \makecell[c]{+ Re-ranking Principles \\ + Cover Image Assessment} 
    & \multicolumn{1}{c}{87.1} 
    & \multicolumn{1}{c}{78.9} 
    & \multicolumn{1}{c}{68.5} 
    & \multicolumn{1}{c}{60.2} 
    & \multicolumn{1}{c}{\underline{64.5}}
    & \multicolumn{1}{c}{87.9} 
    & \multicolumn{1}{c}{80.6} 
    & \multicolumn{1}{c}{68.2} 
    & \multicolumn{1}{c}{62.1} 
    & \multicolumn{1}{c}{\underline{66.5}} 
    & \multicolumn{1}{c}{83.9} 
    & \multicolumn{1}{c}{78.5} 
    & \multicolumn{1}{c}{62.1} 
    & \multicolumn{1}{c}{61.2} 
    & \multicolumn{1}{c}{\underline{63.2}} 
    & \multicolumn{1}{c}{91.6} 
    & \multicolumn{1}{c}{77.9} 
    & \multicolumn{1}{c}{79.8} 
    & \multicolumn{1}{c}{56.7} 
    & \multicolumn{1}{c}{64.8}  \\
    \midrule

    \textbf{ReasonRank} 
    & \multicolumn{1}{c}{69.7} 
    & \multicolumn{1}{c}{67.0} 
    & \multicolumn{1}{c}{54.9} 
    & \multicolumn{1}{c}{52.5} 
    & \multicolumn{1}{c}{55.1}
    & \multicolumn{1}{c}{70.6} 
    & \multicolumn{1}{c}{68.8} 
    & \multicolumn{1}{c}{53.1} 
    & \multicolumn{1}{c}{54.5} 
    & \multicolumn{1}{c}{56.6} 
    & \multicolumn{1}{c}{65.1} 
    & \multicolumn{1}{c}{64.5} 
    & \multicolumn{1}{c}{46.7}
    & \multicolumn{1}{c}{51.3} 
    & \multicolumn{1}{c}{51.1} 
    & \multicolumn{1}{c}{76.1} 
    & \multicolumn{1}{c}{69.3} 
    & \multicolumn{1}{c}{69.9} 
    & \multicolumn{1}{c}{52.6} 
    & \multicolumn{1}{c}{58.9}  \\

 \hdashline

  \makecell[c]{+ Re-ranking Principles}
    & \multicolumn{1}{c}{76.4} 
    & \multicolumn{1}{c}{73.7} 
    & \multicolumn{1}{c}{62.0} 
    & \multicolumn{1}{c}{58.8} 
    & \multicolumn{1}{c}{61.9}
    & \multicolumn{1}{c}{77.1} 
    & \multicolumn{1}{c}{75.1} 
    & \multicolumn{1}{c}{60.4} 
    & \multicolumn{1}{c}{61.2} 
    & \multicolumn{1}{c}{63.1} 
    & \multicolumn{1}{c}{68.8} 
    & \multicolumn{1}{c}{68.4} 
    & \multicolumn{1}{c}{51.1} 
    & \multicolumn{1}{c}{55.3} 
    & \multicolumn{1}{c}{56.1} 
    & \multicolumn{1}{c}{87.8} 
    & \multicolumn{1}{c}{80.9} 
    & \multicolumn{1}{c}{80.9} 
    & \multicolumn{1}{c}{62.2} 
    & \multicolumn{1}{c}{70.0}  \\

 \hdashline

  \makecell[c]{+ Re-ranking Principles \\ + Cover Image Assessment}
    & \multicolumn{1}{c}{78.5} 
    & \multicolumn{1}{c}{75.7} 
    & \multicolumn{1}{c}{65.1} 
    & \multicolumn{1}{c}{60.8} 
    & \multicolumn{1}{c}{64.3}
    & \multicolumn{1}{c}{78.6} 
    & \multicolumn{1}{c}{76.1} 
    & \multicolumn{1}{c}{63.4} 
    & \multicolumn{1}{c}{61.3} 
    & \multicolumn{1}{c}{65.2} 
    & \multicolumn{1}{c}{71.2} 
    & \multicolumn{1}{c}{70.6} 
    & \multicolumn{1}{c}{54.6} 
    & \multicolumn{1}{c}{57.5} 
    & \multicolumn{1}{c}{58.5} 
    & \multicolumn{1}{c}{90.2} 
    & \multicolumn{1}{c}{\underline{83.7}} 
    & \multicolumn{1}{c}{83.7} 
    & \multicolumn{1}{c}{\underline{65.6}} 
    & \multicolumn{1}{c}{\underline{72.9}}  \\

    \midrule

    \textbf{ReaRank} 
    & \multicolumn{1}{c}{83.8} 
    & \multicolumn{1}{c}{75.2} 
    & \multicolumn{1}{c}{65.3} 
    & \multicolumn{1}{c}{56.4} 
    & \multicolumn{1}{c}{57.2}
    & \multicolumn{1}{c}{83.6} 
    & \multicolumn{1}{c}{76.8}
    & \multicolumn{1}{c}{62.3} 
    & \multicolumn{1}{c}{59.0} 
    & \multicolumn{1}{c}{57.9} 
    & \multicolumn{1}{c}{79.8}
    & \multicolumn{1}{c}{75.3} 
    & \multicolumn{1}{c}{56.7} 
    & \multicolumn{1}{c}{57.6} 
    & \multicolumn{1}{c}{56.8} 
    & \multicolumn{1}{c}{90.5} 
    & \multicolumn{1}{c}{73.4} 
    & \multicolumn{1}{c}{82.2} 
    & \multicolumn{1}{c}{52.1} 
    & \multicolumn{1}{c}{57.0}  \\

 \hdashline
 
  \makecell[c]{+ Re-ranking Principles} 
    & \multicolumn{1}{c}{87.2} 
    & \multicolumn{1}{c}{77.6} 
    & \multicolumn{1}{c}{70.0} 
    & \multicolumn{1}{c}{59.2} 
    & \multicolumn{1}{c}{60.5}
    & \multicolumn{1}{c}{88.3} 
    & \multicolumn{1}{c}{80.0} 
    & \multicolumn{1}{c}{69.6} 
    & \multicolumn{1}{c}{62.1} 
    & \multicolumn{1}{c}{63.1} 
    & \multicolumn{1}{c}{83.0} 
    & \multicolumn{1}{c}{77.9} 
    & \multicolumn{1}{c}{60.8} 
    & \multicolumn{1}{c}{61.0} 
    & \multicolumn{1}{c}{58.8} 
    & \multicolumn{1}{c}{92.8} 
    & \multicolumn{1}{c}{74.8}
    & \multicolumn{1}{c}{85.1} 
    & \multicolumn{1}{c}{53.3} 
    & \multicolumn{1}{c}{60.7}  \\

 \hdashline

  \makecell[c]{+ Re-ranking Principles \\ + Cover Image Assessment} 
    & \multicolumn{1}{c}{\underline{88.5}} 
    & \multicolumn{1}{c}{\underline{79.6}} 
    & \multicolumn{1}{c}{\underline{72.4}} 
    & \multicolumn{1}{c}{\underline{61.1}} 
    & \multicolumn{1}{c}{63.2}
    & \multicolumn{1}{c}{\underline{89.1}} 
    & \multicolumn{1}{c}{\underline{81.5}} 
    & \multicolumn{1}{c}{\underline{71.0}} 
    & \multicolumn{1}{c}{\underline{63.3}} 
    & \multicolumn{1}{c}{65.1} 
    & \multicolumn{1}{c}{\underline{85.1}} 
    & \multicolumn{1}{c}{\underline{80.3}} 
    & \multicolumn{1}{c}{\underline{65.0}} 
    & \multicolumn{1}{c}{\underline{63.3}} 
    & \multicolumn{1}{c}{62.4} 
    & \multicolumn{1}{c}{\underline{93.1}} 
    & \multicolumn{1}{c}{76.6} 
    & \multicolumn{1}{c}{\underline{85.6}} 
    & \multicolumn{1}{c}{55.3} 
    & \multicolumn{1}{c}{62.6}  \\
    
    \midrule

    \textbf{\MethodName} 
    & \multicolumn{1}{c}{\textbf{97.7}} 
    & \multicolumn{1}{c}{\textbf{96.7}} 
    & \multicolumn{1}{c}{\textbf{89.0}} 
    & \multicolumn{1}{c}{\textbf{88.1}} 
    & \multicolumn{1}{c}{\textbf{88.6}}
    & \multicolumn{1}{c}{\textbf{97.4}} 
    & \multicolumn{1}{c}{\textbf{96.7}} 
    & \multicolumn{1}{c}{\textbf{86.9}} 
    & \multicolumn{1}{c}{\textbf{89.3}} 
    & \multicolumn{1}{c}{\textbf{88.8}} 
    & \multicolumn{1}{c}{\textbf{97.4}} 
    & \multicolumn{1}{c}{\textbf{96.7}} 
    & \multicolumn{1}{c}{\textbf{86.5}} 
    & \multicolumn{1}{c}{\textbf{88.0}} 
    & \multicolumn{1}{c}{\textbf{87.7}} 
    & \multicolumn{1}{c}{\textbf{98.5}} 
    & \multicolumn{1}{c}{\textbf{96.9}} 
    & \multicolumn{1}{c}{\textbf{95.0}} 
    & \multicolumn{1}{c}{\textbf{87.2}} 
    & \multicolumn{1}{c}{\textbf{89.8}}  \\

    \bottomrule
    \end{tabular}}

  \label{exp_main}%

\end{table*}%

\subsection{Experimental setup}

\subsubsection{Evaluation Datasets}
To evaluate the performance of the Rich-Media Re-Ranker in a real-world search system, we constructed an industrial dataset by sampling diverse user queries from online logs, covering daily-life questions, travel planning, comparative inquiries, and so on. For each query, we apply the Query Planner to form the final re-ranking candidate set. The dataset contains 14,115 queries in total, consisting of 3,582 complex queries, 6,859 broad-needs queries, and 3,674 simple queries, with each query associated with an average of 20 candidates.

\subsubsection{Baselines and Evaluation Metrics}

We compare the proposed Rich-Media Re-Ranker against several representative LLM-based re-ranking methods, including ReaRank~\cite{zhang2025rearank}, Rank-R1~\cite{zhuang2025rank}, and ReasonRank~\cite{liu2025reasonrank}. To comprehensively evaluate performance, we adopt top-K Normalized Discounted Cumulative Gain (N@K) and Recall (R@K) to measure the performance of varying browsing depths, and RBO to assess overall ranking.

\subsection{Performance in Rich-Media Search}
We evaluate the proposed rich-media re-ranker against state-of-the-art baselines on our constructed rich-media search dataset. For a fair comparison, we additionally enhance the baseline methods by incorporating the comprehensive re-ranking principle (including side information) and the visual signals extracted by our VLM evaluator. The main results are presented in Table~\ref{exp_main}, from which we draw the following conclusions:
\begin{itemize}[leftmargin=0.3cm]
    \item The Rich-Media Re-Ranker achieves the best performance in all categories of queries, outperforming the runner-up methods by significant margins of 17.1\% in NDCG@10 and 27.0\% in Recall@10 overall. This superiority can be attributed to two primary factors: (1) Firstly, baseline methods rely solely on relevance as the ranking criterion. In contrast, our reranker comprehensively considers relevance, demand satisfaction, information gain, novelty, and visual perception. This holistic approach aligns better with real-world user needs and constitutes a more practical improvement. (2) Secondly, baseline methods depend exclusively on textual content for ranking, failing to fully exploit the abundant visual information inherent in the rich-media scenario. In contrast, our approach leverages a post-trained VLM evaluator to extract refined visual signals from cover images.
    \item To further investigate the contributing factors, we incrementally enhanced the baseline methods by integrating our designed re-ranking principle and the extracted visual cover signals. The results show a progressive improvement in their performance, thereby supporting the validity of our design.
    \item Despite the observed enhancement from incorporating our components, the augmented baselines still fall short of our performance. This performance gap is primarily due to their not undergoing sufficient post-training within the corresponding re-ranking paradigm, leading to inadequate adaptation to the specific demands of the rich-media re-ranking scenario. This observation is further supported by the comparable performance on simple queries, contrasted with the pronounced gaps observed for complex and broad-needs queries.
\end{itemize}

\begin{figure*}[t]
    \centering
        \begin{tabular}{@{}c@{\hspace{0.005\textwidth}}c@{\hspace{0.005\textwidth}}c@{}}
        \begin{tabular}{@{}c@{\hspace{0.005\textwidth}}c@{}}
            \includegraphics[width=0.33\textwidth]{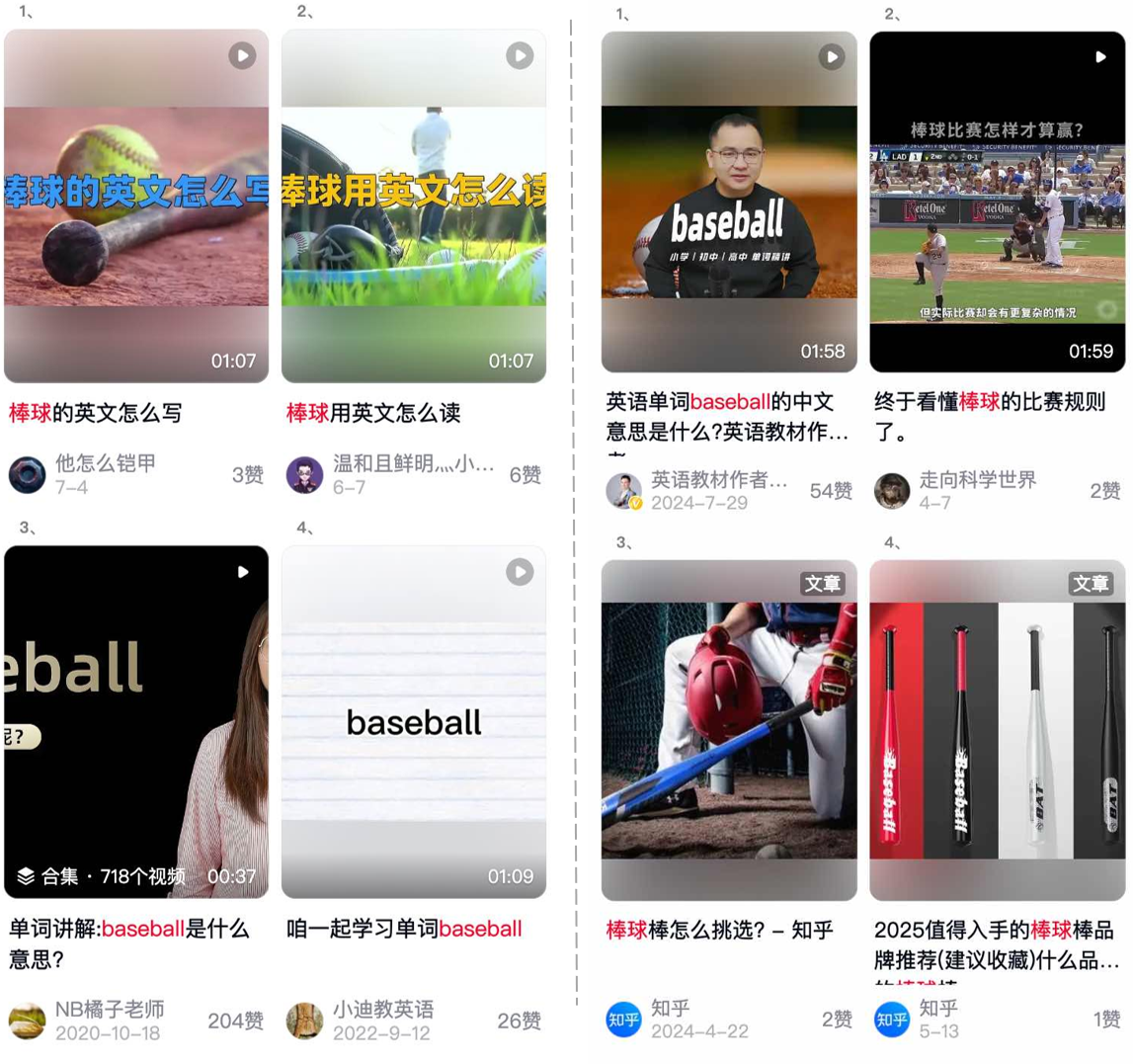} \\
            \multicolumn{2}{c}{\footnotesize\makecell{\textbf{(a)} Query="baseball" \\ }}
        \end{tabular}
 &
        \begin{tabular}{@{}c@{\hspace{0.005\textwidth}}c@{}}
            \includegraphics[width=0.33\textwidth]{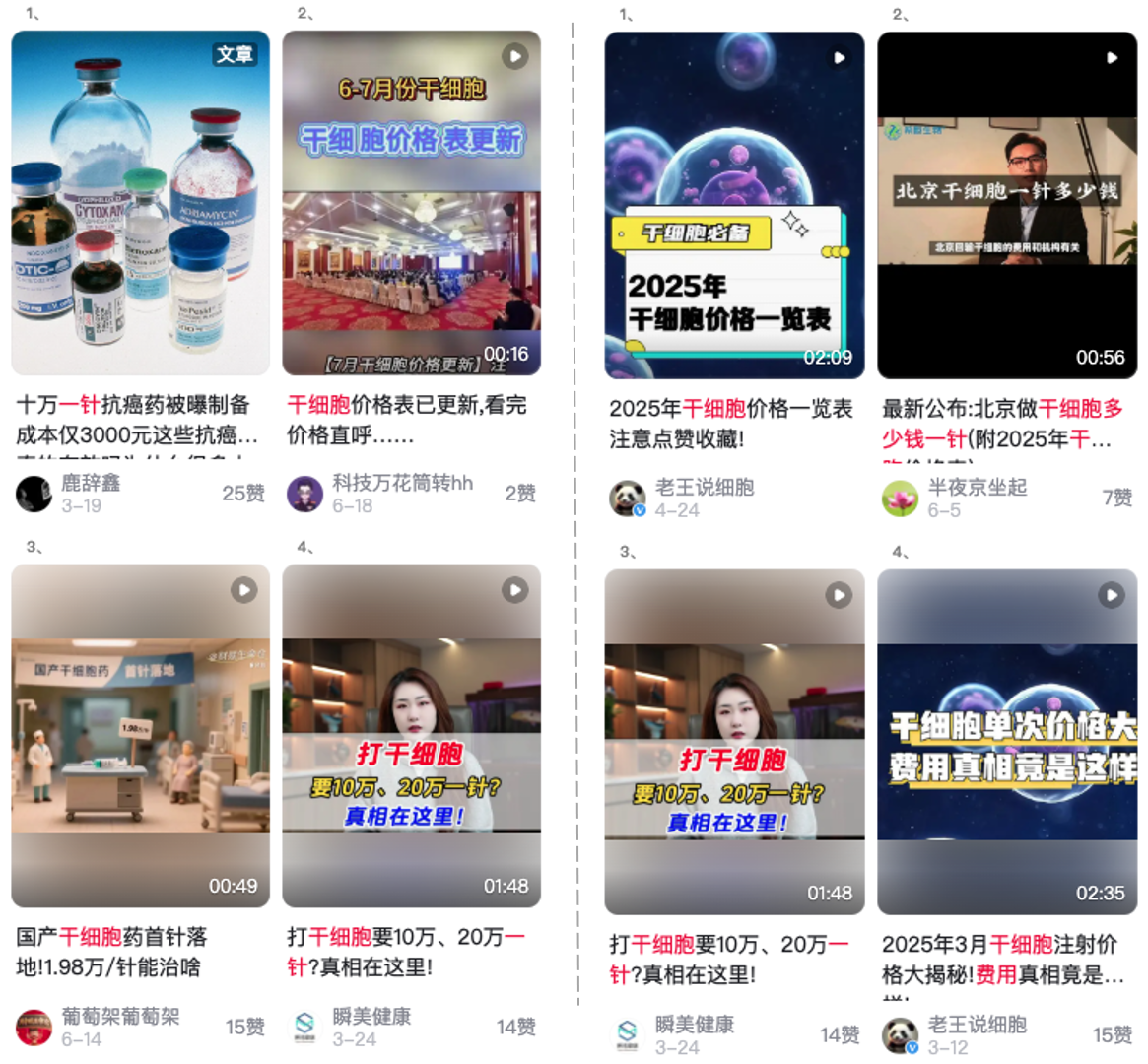} \\
            \multicolumn{2}{c}{\footnotesize\makecell{\textbf{(b)} Query="How much does one injection of \\ stem cells cost?"}}
        \end{tabular}
 &
        \begin{tabular}{@{}c@{\hspace{0.005\textwidth}}c@{}}
            \includegraphics[width=0.33\textwidth]{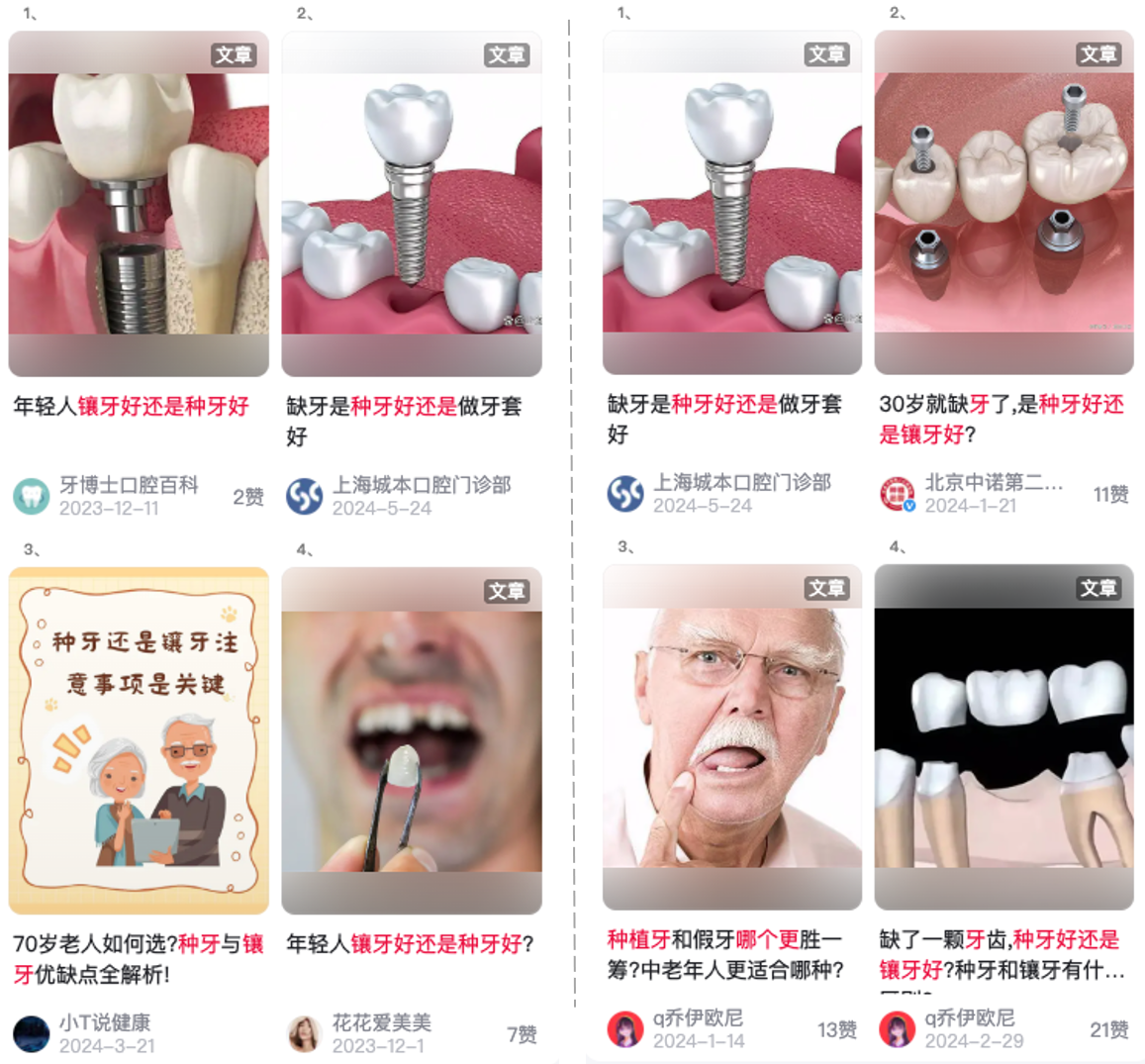} \\
            \multicolumn{2}{c}{\footnotesize\makecell{\textbf{(c)} Query="Which is healthier, dental implants \\ or dentures?"}}
        \end{tabular}
    \end{tabular}
    \captionsetup{skip=5pt}
    \caption{Case study on online rich-media search system. Left: Online System, Right: Our Rich Media Re-ranker.}
    \label{case_study}
\end{figure*}

\setlength{\equalcolwidth}{0.045\textwidth} 

\begin{table}[t]
\renewcommand{\arraystretch}{1.1}
  \centering
  \captionsetup{skip=5pt}
  \caption{The ablation study.}  
    \resizebox{1.0\linewidth}{!}{
    \setlength{\tabcolsep}{2.1pt}
    \begin{tabular}{@{}cc*{6}{>{\centering\arraybackslash}p{\equalcolwidth}}@{}}
    \toprule
     \textbf{Method} & N@4  & N@10  & R@4 & R@10 & RBO\\
    \midrule

    \textbf{\MethodName}  & \textbf{97.7} & \textbf{96.7} & \textbf{89.0} & \textbf{88.1} & \textbf{88.6} \\

    \midrule
 
  \makecell[c]{\textit{w/o} Re-ranking Principle} 
    & \multicolumn{1}{c}{89.8} 
    & \multicolumn{1}{c}{87.6} 
    & \multicolumn{1}{c}{73.7} 
    & \multicolumn{1}{c}{72.2} 
    & \multicolumn{1}{c}{74.0} \\

  \makecell[c]{\textit{w/o} Cover Image Assessment} 
    & \multicolumn{1}{c}{96.0} 
    & \multicolumn{1}{c}{94.3} 
    & \multicolumn{1}{c}{83.6} 
    & \multicolumn{1}{c}{82.6} 
    & \multicolumn{1}{c}{83.1} \\

  \makecell[c]{\textit{w/o} VLM Post-Train} 
    & \multicolumn{1}{c}{96.3} 
    & \multicolumn{1}{c}{94.4} 
    & \multicolumn{1}{c}{83.9} 
    & \multicolumn{1}{c}{83.3} 
    & \multicolumn{1}{c}{84.7} \\

  \makecell[c]{\textit{w/o} LLM Post-Train}
    & \multicolumn{1}{c}{85.6} 
    & \multicolumn{1}{c}{70.1} 
    & \multicolumn{1}{c}{70.6} 
    & \multicolumn{1}{c}{52.6} 
    & \multicolumn{1}{c}{57.2} \\

    \bottomrule
    \end{tabular}}

  \label{exp_ablation}%

\end{table}%

\subsection{Ablation Studies}

To further investigate, we conducted ablation studies on several variants of our method:
\begin{itemize}[leftmargin=0.3cm]
    \item \textit{w/o} Re-ranking principle: We replaced the designed re-ranking principle with a simple relevance-based principle, while only retaining the title and content of candidate results.
    \item \textit{w/o} Cover Image Assessment: We removed all cover value signals of candidate results.
    \item \textit{w/o} LLM Post-Train: We used the LLM without post-training as the re-ranker.
    \item \textit{w/o} VLM Post-Train: We replaced all cover value signals with outputs from the VLM without post-training.
\end{itemize}

Results are summarized in Table~\ref{exp_ablation}. First, we observe that removing any key component consistently degraded performance, highlighting the importance of each part. Specifically, (1) For \textit{w/o} Re-ranking principle, by focusing only on text relevance between candidates and the query during re-ranking, the model ignores other user satisfaction-critical factors such as demand fulfillment, information gain, novelty, and visual perception. Though the re-ranked results are reasonable, they fail to fully benefit the overall search experience. (2) For \textit{w/o} Cover Image Assessment, despite achieving competitive re-ranking results using textual content alone, the absence of visual signals prevents necessary adjustments based on visual perception in certain cases. (3) For \textit{w/o} VLM Post-Train, the VLM without specialized post-training exhibits poor classification accuracy in cover image relevance and quality dimensions, rendering the provided visual signals unreliable. (4) For \textit{w/o} LLM Post-Train, despite having detailed re-ranking principles and candidate information, the zero-shot LLM struggles to effectively execute the complex re-ranking task.

\begin{table}[t]
\renewcommand{\arraystretch}{1.1}
  \centering
  \captionsetup{skip=5pt}
  \caption{The results of the online A/B test. All the performances gains are statistically significant at $p < 0.05$.}  
    \resizebox{0.9\linewidth}{!}{
    \setlength{\tabcolsep}{2.1pt}

    \begin{tabular}{@{} >{\centering\arraybackslash}p{4cm} >{\centering\arraybackslash}p{3cm} @{}}
    \toprule

     \textbf{Online Metrics} & \textbf{Improv.} \\
    \midrule
    Overall Click Rate & + 0.34\% \\
    Positive Interaction Ratio & + 0.33\% \\
    Satisfaction Rate & + 0.43\% \\
    
    \bottomrule

    \end{tabular}

    }

  \label{exp_ab_test}%

\end{table}%

\section{Online Applications}

\subsection{Performance of A/B Test}

To evaluate the real-world performance of our method, we deployed it in the industrial rich-media search system for the A/B test. We randomly select 0.5\% of online traffic as the experimental group for Rich-Media Re-Ranker, while another 0.5\% of traffic served as the baseline group. The experiment lasted for 3 days. As shown in Table~\ref{exp_ab_test}, we achieve significant improvements across all metrics. The increased Overall Click Rate and Positive Interaction Ratio reflect enhanced coverage of search intents that better align with users' information needs and content preferences. The higher Satisfaction Rate directly captures the enhanced search experience. These gains validate the alignment with user-centric objectives.

\subsection{Case Study}

Figure~\ref{case_study} illustrates three real cases collected from the online rich media search system. For each group, the left side displays the ranking presented to users under the original sorting strategy, while the right side shows the results re-ranked by our method:
\begin{itemize}[leftmargin=0.3cm]
    \item \textbf{Case (a)}: Original results exhibited severe homogeneity, exclusively focusing on English definitions of "baseball." In contrast, re-ranked results accounted for multidimensional potential user needs, not only addressing the likely primary need for English definitions but also covering secondary needs such as baseball game rules and equipment guidance.
    \item \textbf{Case (b)}: In the original ranking, the first item’s title was perceptually irrelevant to the query, and the third item’s low-value cover image failed to convey useful information. The re-ranked results demonstrated superior performance in both content relevance and cover image value.
    \item \textbf{Case (c)}: Though both original and re-ranked results addressed the user’s core query, the re-ranked version further provides a comparison between dental implants and dentures, offering a richer selection for decision-making.
\end{itemize}


\section{Conclusion}
In this paper, we propose Rich-Media Re-Ranker, a novel re-ranking framework addressing two critical limitations in search systems: inadequate modeling of multifaceted user intents and neglect of rich side information, including visual characteristics. Firstly, our session-aware Query Planner decomposes user queries into complementary sub-intents based on session context for comprehensive user need coverage. Then, a VLM-based Cover Image Value Assessment module explicitly evaluates visual relevance and quality. Additionally, the primary content of candidates and comprehensive side information, including cover image signals, is integrated by an LLM-based re-ranker, which holistically evaluates content relevance, quality, information novelty, information gains, visual perception, and so on for interpretable re-ranking. Finally, these components are enhanced via multi-task reinforcement learning. Extensive experiments demonstrate that the framework significantly outperforms state-of-the-art baselines, while its successful deployment in a large-scale search system validates substantial improvements in user engagement and satisfaction metrics.

\begin{acks}
This work is supported by the National Natural Science Foundation of China through grants No.62302023 and No.62225202, and by the Fundamental Research Funds for the Central Universities. We owe sincere thanks to all authors for their valuable efforts and contributions.
\end{acks}

\setcounter{figure}{0}
\setcounter{table}{0}
\setcounter{equation}{0}
\setcounter{section}{0}
\renewcommand\thesection{\Alph{section}}
\renewcommand\thefigure{\thesection.\arabic{figure}}
\renewcommand\thetable{\thesection.\arabic{table}}
\renewcommand\theequation{\thesection.\arabic{equation}}

\section{Further Details on Online Deployment}
Due to the scale of model parameters and limited resources, we adopt the T+1 activation approach. Specifically, for a given day's user queries, we perform offline inference to obtain the re-ranked results, which are then written into the cache for retrieval on the following day. Consequently, our active traffic does not include personalized serving routes.

\section{Implementation Details}

\subsection{Query Planner}
\label{sec:prompt_planner}

The prompt for the Query Planner is designed to systematically decompose the user query based on session context for comprehensive intent coverage. We first classify queries into three types: complex query, broad-needs query, or simple query. For complex queries, we instruct the model to break down comparative or multifaceted aspects into independent sub-queries inferred from session history. For broad-needs queries, we guide it to extend coverage by generating non-redundant sub-queries across diverse dimensions derived from contextual intent. We further direct the assignment of intent dimensions like High Freshness for time sensitivity or Authoritativeness for credibility to each sub-query, enabling alignment with decomposed user requirements.

\begin{tcolorbox}[breakable,title={Query Planning Task for Query Planner}]

As a query planning expert, you need to first classify user query and
then decompose user query based on session context to capture
multifaceted intents. Please think step by step before answering, placing
your reasoning within <thinking></thinking> tags, and provide the final
answer within <answer></answer> tags.

\smallskip

I will provide a user query and corresponding session context. Please
follow the following instructions:

\smallskip

\textbf{Step 1: Classify the user query}

\begin{itemize}[leftmargin=0.3cm]
    \item Complex Query: Requires comparisons, reasoning, calculations, or multi-step information integration to reach a conclusion.
    \item Broad-needs Query: Broad/vague/incomplete intent with potential multidimensional needs across different directions.
    \item Simple Query: Explicit, singular intent that can be directly answered.
\end{itemize}

\smallskip

\textbf{Step 2: Decompose the original query based on session context}

\begin{itemize}[leftmargin=0.3cm]
    \item For Complex Query: Identify portions requiring comparisons or multi-step information gathering, breaking them down into independently searchable sub-queries inferred from session context.
    \item For Broad-needs Query: Recognize and extend possible multidimensional aspects using inferred session intent, generating sub-queries that broadly yet non-redundantly cover potential intents.
    \item For Simple Query: Rewriting it appropriately if obvious typo error is detected; otherwise, preserve the original query.
\end{itemize}

\smallskip

\textbf{Step 3: Assign intent dimensions}

Assign one or more dimensions to each sub-query:

\begin{itemize}[leftmargin=0.3cm]
    \item High Freshness
    \item Authoritativeness
    \item Personal Experience
\end{itemize}

\smallskip

\textbf{User Query}: ...

\smallskip

\textbf{Session Context}: ...

The final answer needs to be provided in JSON format, as follows:

\{

    \quad "query\_type": "Complex Query/Broad-needs Query/Simple Query",
    
    \quad "sub\_queries": [
    
        \quad \quad \{"sub\_query": "Sub-query 1", "intent\_dimensions": ["Dimension1", "Dimension2"]\},
        
        \quad \quad \{"sub\_query": "Sub-query 2", "intent\_dimensions": ["Dimension1"]\}
        
    \quad ]
    
\}
\end{tcolorbox}

\subsection{LLM Re-ranker}
\label{sec:prompt_rerank}
The LLM Re-ranker prompt is designed to optimize search rankings by prioritizing candidates that directly align with the core intent of the user query while providing comprehensive overviews. We then supplement these primary results with diverse perspectives or novel angles to maximize information gain and avoid redundancy. When candidates exhibit comparable informational value, we instruct the model to break ties by favoring those with higher cover image relevance, superior cover image quality, or stronger user engagement signals such as click-through and content completion rates. We further ensure candidates misaligned with their designated intent dimension are deprioritized. This integrated approach balances core relevance, information novelty, visual appeal, and intent alignment to enhance user satisfaction.

\begin{tcolorbox}[breakable,title={Re-ranking Task for LLM Re-ranker}]
As a re-ranking expert, you need to re-rank the candidate result list based on the user query to maximize user search experience. Please think step by step before answering, placing your reasoning within <thinking></thinking> tags, and provide the final answer within <answer></answer> tags.

\smallskip

I will provide a user query and a corresponding list of candidate search results. Each candidate result includes ID, Title, Content, Intent Dimension, Publish Time, Cover Image Relevance, Cover Image Quality, Click-Through Rate, and Content Completion Rate. \textbf{Please follow the following principles for re-ranking}:

\begin{itemize}[leftmargin=0.3cm]
    \item Prioritize results highly relevant to the user query that provide an overview or cover core themes.
    \item Subsequently supplement with results from different perspectives, details, or novel angles. Ensure each new result offers significant information gain while avoiding redundancy.
    \item When information value is comparable, prioritize results with higher cover value, higher content quality, or higher click-through and content completion rates.
    \item For each candidate, reduce consideration if their content does not align with the corresponding Intent Dimension.
\end{itemize}

\smallskip

\textbf{User Query}: ...

\smallskip

\textbf{Candidate results are as follows}:

[ID] 1

[Title] ...

[Content] ...

[Intent Dimension] ...

[Publish Time] ...

[Cover Image Relevant] ...

[Cover Image Quality] ...

[Click-Through Rate] ...

[Content Completion Rate] ...

\smallskip

[ID] 2

......

\smallskip

The final answer must contain only a ID array that has been re-ranked,
including all original IDs without exceeding candidate indices. Format
example: [3,1,5,2,4].

\end{tcolorbox}

\subsection{VLM Evaluator}
\label{sec:prompt_vlm}
\subsubsection{Relevance assessment}

The VLM Evaluator's relevance assessment prompt is designed to assign integer ratings from 1 to 4 based on alignment between the cover image and the user query. We define four relevance tiers: strongly relevant images fully satisfy the core query intent, relevant images address secondary aspects or provide general reference, weakly relevant images offer minimal reference value without fulfilling primary needs, and irrelevant images show no meaningful connection.

\begin{tcolorbox}[breakable,title={Cover Image Relevance Assessment Task for VLM Evaluator}]
As a cover image relevance assessment expert, you need to evaluate and grade the relevance of cover image to the user query. Please think step by step before answering, placing your reasoning within <thinking></thinking> tags, and provide the final answer within <answer></answer> tags.

\smallskip

Assign an integer rating from [1,2,3,4] representing increasing relevance levels, specifically:

\begin{itemize}[leftmargin=0.3cm]
    \item \textbf{Rating 4}: Strongly Relevant – The image theme is highly aligned with the query and can fully satisfy the user's needs.
    \item \textbf{Rating 3}: Relevant – Generally usable, can serve as a reference, or addresses secondary aspects of the query.
    \item \textbf{Rating 2}: Weakly Relevant – Doesn't fulfill core needs but offers slight reference.
    \item \textbf{Rating 1}: Irrelevant.
\end{itemize}

\smallskip

\textbf{User Query}: ...

\end{tcolorbox}

\subsubsection{Quality assessment}

The VLM Evaluator's quality assessment prompt is designed to assign integer ratings from 1 to 4 through the comprehensive quality analysis of the cover image. We define four quality tiers: high-quality images exhibit strong aesthetic appeal or rich informational value, average images show no objective flaws, poor images contain minor issues like slight blurring without severely hindering comprehension, and very poor images display severe flaws such as heavy watermarks or extreme distortion.

\begin{tcolorbox}[breakable,title={Cover Image Quality Assessment Task for VLM Evaluator}]
As a cover image quality assessment expert, you need to evaluate and grade the quality of cover image. Please think step by step before answering, placing your reasoning within <thinking></thinking> tags, and provide the final answer within <answer></answer> tags.

\smallskip

Conduct a comprehensive analysis across multiple dimensions including aesthetic composition, color vibrancy, layout harmony, and the absence of disruptive elements (such as watermarks or blurring). 

\smallskip

Assign an integer rating from [1,2,3,4] representing increasing quality levels, specifically:
\begin{itemize}[leftmargin=0.3cm]
    \item \textbf{Rating 4}: High Quality – Visually appealing images that align with popular aesthetic standards, or possess rich informational value.
    \item \textbf{Rating 3}: Average – Images with no objective quality flaws.
    \item \textbf{Rating 2}: Poor – Minor objective issues that don't severely hinder information comprehension (e.g., slight blurring, black borders, or mild stretching).
    \item \textbf{Rating 1}: Very Poor – Marketing advertisements, or severe objective flaws like blurred subjects, poor visuals, extreme cropping, or prominent watermarks (e.g., heavy blurring, cropped-out key elements, or severe distortion).
\end{itemize}

\end{tcolorbox}

\section{Experiment Details}

\subsection{Datasets}

For the re-ranking dataset, we sample 1,411 queries (10\%) for comparative evaluation, with the remainder used for model training. The labels adopted for model training are generated by DeepSeek-R1, while evaluation labels follow far stricter criteria and predominantly depend on human judgment, which eliminates the risk of evaluation bias. For the cover image value assessment dataset, each grade of the two tasks contains 500 images for training and 100 images for evaluation, and all data labels in this dataset are high-quality manually annotated results.

\subsection{Details about Post-Training}
Our VLM evaluator employs Qwen-2.5 7B as its backbone model, undergoing post-training via the VLM-R1 framework$^1$. Simultaneously, the LLM reranker utilizes Qwen3 4B as its backbone model, with post-training conducted through the VERL framework$^2$. For both, full-parameter training was performed on 8×A800 GPUs.

\def\thefootnote{$^1$}\footnotetext{https://github.com/om-ai-lab/VLM-R1}
\def\thefootnote{$^2$}\footnotetext{https://github.com/volcengine/verl}

\newpage
\clearpage

\section*{GenAI Usage Disclosure}
The use of GenAI tools solely for linguistic polishing, grammar correction, and expression refinement of the manuscript text.

\bibliographystyle{ACM-Reference-Format}
\bibliography{ref}

\end{document}